\documentclass[twocolumn,floatfix,nofootinbib,preprintnumbers,eqsecnum]{revtex4}
\usepackage{dcolumn}%
\usepackage{bm}%
\usepackage{fullpage}
\usepackage[centertags]{amsmath}
\allowdisplaybreaks[1]
\usepackage{amsbsy}
\usepackage{amsfonts}
\usepackage{amssymb}
\usepackage{eufrak}
\usepackage[dvips]{graphicx}
\begin{document}
\preprint{hep-ph/0209103}
\title{Current Status of Neutrino Masses and Mixings}
\author{Carlo Giunti}
\email{giunti@to.infn.it}
\homepage{http://www.to.infn.it/~giunti}
\affiliation{
INFN, Sezione di Torino, and Dipartimento di Fisica Teorica, Universit\`a di Torino,
Via P. Giuria 1, I--10125 Torino, Italy
}
\date{10 September 2002}
\begin{abstract}
The evidences in favor of solar and atmospheric neutrino
oscillations are briefly reviewed
and
shown to be gracefully
accommodated in the framework of
three-neutrino mixing with bilarge mixing.
\textsf{
Talk presented at the
31$^{st}$ International Conference on High Energy Physics
``ICHEP02'', 24--31 July 2002, Amsterdam.
}
\end{abstract}
\maketitle

\section{Introduction}
\label{Introduction}

Neutrino physicists are living a very interesting time,
in which several fundamental facts,
\renewcommand{\labelenumi}{\theenumi}
\renewcommand{\theenumi}{(F\arabic{enumi})}
\begin{enumerate}
\item
\label{F-three-active}
there are three active light neutrinos
($\nu_e$, $\nu_\mu$, $\nu_\tau$),
\item
\label{F-mass-limit}
the mass of light neutrinos is smaller than a few eV,
\item
\label{F-sun}
there are solar
$\nu_e \to \nu_\mu, \nu_\tau$
transitions,
\item
\label{F-atm}
there are atmospheric
$\nu_\mu \to \nu_\tau$
transitions,
\item
\label{F-nu-e}
atmospheric and long-baseline
$\nu_e$
oscillations are suppressed,
\item
\label{F-sbl}
short-baseline oscillations of
$\nu_e$
and
$\nu_\mu$
are suppressed,
\end{enumerate}
have been experimentally established
with a reasonable degree of reliability
and several fundamental mysteries,
\renewcommand{\labelenumi}{\theenumi}
\renewcommand{\theenumi}{(M\arabic{enumi})}
\begin{enumerate}
\item
\label{M-absolute-scale}
which is the absolute scale of neutrino masses?
\item
\label{M-nature}
are neutrinos Dirac or Majorana particles?
\item
\label{M-number}
which is the number of massive neutrinos?
\item
\label{M-sterile}
are there transitions of active neutrinos into sterile states?
\item
\label{M-CP}
is there CP violation in the lepton sector?
\end{enumerate}
are under investigation.
The known facts about
neutrino masses and mixings
are important for the formulation of theoretical frameworks
and of experimental strategies for the future
clarification of the mysteries.
For an extensive and updated list of references
see Ref.~\cite{Neutrino-Unbound}.

In this review,
I will briefly summarize the experimental evidences in favor
of the above-mentioned facts (\ref{F-mass-limit})--(\ref{F-nu-e})
and I will show that they can be accommodated
elegantly in the framework of three-neutrino mixing.
I will not discuss the already well known
experimental proof that there are only
three active light neutrinos (fact~\ref{F-three-active} above,
see Ref.~\cite{PDG}).
For lack of space
I will not review the abundant experimental evidences
of the fact~\ref{F-sbl} above
(see Refs.~\cite{Neutrino-Unbound,PDG,BGG-review-98-brief})
and
I will not discuss the status of the four-neutrino mixing hypothesis
(see Ref.~\cite{Neutrino-Unbound,BGG-review-98-brief}),
which was proposed in order to accommodate also the claim of
short-baseline
$ \bar\nu_\mu \to \bar\nu_e $
transitions in the LSND experiment
(see Ref.~\cite{Church:2002tc}).
This hypothesis is now disfavored by data
(see Ref.~\cite{hep-ph/0207157})
and my opinion is that it should be put under
hibernation waiting for its
final elimination or resurrection
by new data
(especially crucial will be the check of
the LSND result in the MiniBooNE experiment
\cite{MiniBooNE-Nu2002-brief}).

\section{Solar Neutrinos}
\label{Solar Neutrinos}

All solar neutrino experiments
(Homestake \cite{Cleveland:1998nv},
SAGE \cite{astro-ph/0204245},
GALLEX \cite{Hampel:1998xg},
GNO \cite{Altmann:2000ft},
Super-Kamiokande \cite{Fukuda:2002pe},
SNO \cite{Ahmad:2002jz})
measured a suppression of the
observed rate
with respect to the one predicted by the
Standard Solar Model (SSM) \cite{BP2000-brief}.
The only exception is
the neutral-current reaction
$ \nu + d \to p + n + \nu $
in the SNO experiment \cite{Ahmad:2002jz},
which is equally sensitive to all active neutrino flavors,
whose rate is in good agreement
with the SSM.
This measurement has marked the triumph of the
SNO experiment,
which measured also a suppression of a factor
$0.35 \pm 0.02$
of the flux of solar $\nu_e$'s
with respect to the SSM
through the charged-current reaction
$ \nu_e + d \to p + p + e^- $ \cite{Ahmad:2002jz}.
The discrepancy between the neutral-current and charged-current
suppressions
is a clear and impressive model-independent evidence
of transitions of solar $\nu_e$'s
into active $\nu_\mu$ or $\nu_\tau$
(fact~\ref{F-sun} above).
The existence of solar
$ \nu_e \to \nu_\mu,\nu_\tau $
transitions
is also supported
by the comparison of the
different suppressions of the rate
of the SNO charged-current reaction
\cite{Ahmad:2002jz}
and the rate of the
elastic scattering reaction
$ \nu + e^- \to \nu + e^- $,
which is mainly sensitive to $\nu_e$
but has also a small sensitivity to
$\nu_\mu$ and $\nu_\tau$
($
\sigma^{\mathrm{ES}}_{\nu_\mu,\nu_\tau}
\simeq
\sigma^{\mathrm{ES}}_{\nu_e}/6
$),
measured in Super-Kamiokande
\cite{Fukuda:2002pe}
and SNO \cite{Ahmad:2002jz}.

Let me emphasize also the importance of
the Homestake experiment \cite{Cleveland:1998nv},
in which the solar neutrino problem was discovered,
and the Gallium
SAGE \cite{astro-ph/0204245},
GALLEX \cite{Hampel:1998xg}
and
GNO \cite{Altmann:2000ft} experiments,
which are sensitive to the fundamental flux of $pp$
neutrinos
(see Ref.~\cite{BP2000-brief}).
Moreover,
the results of all neutrino experiments
are necessary in order to get information
on the neutrino mixing parameters.

The global analysis of all solar neutrino data
in terms of
$\nu_e \to \nu_\mu,\nu_\tau$
performed in Ref.~\cite{Bahcall:2002hv}
yielded
\begin{equation}
\begin{array}{l}
0.24
<
\tan^2 \vartheta_{\mathrm{S}}
<
0.89
\,,
\\
2.3 \times 10^{-5}
<
\Delta{m}^2_{\mathrm{S}} / \mathrm{eV}^2
<
3.7 \times 10^{-4}
\,,
\end{array}
\label{LMA}
\end{equation}
at 99.73\% C.L.
($3\sigma$),
where
$\Delta{m}^2_{\mathrm{S}}$
is the relevant neutrino squared-mass difference
and
$\vartheta_{\mathrm{S}}$
is the effective mixing angle
in two-generation analyses
of solar neutrino data.
In Eq.~(\ref{LMA})
I reported only the boundaries of the
so-called LMA region
(see Ref.~\cite{BGG-review-98-brief}),
which is
currently favored,
because it is much larger than other regions
(a LOW region and three VAC regions appear at 99\% C.L.)
and
it contains the minimum of the $\chi^2$
(best fit)
at
\begin{equation}
\tan^2 \vartheta_{\mathrm{S}}
\simeq
0.42
\,,
\
\Delta{m}^2_{\mathrm{S}}
\simeq
5.0 \times 10^{-5} \,\mathrm{eV}^2
\,.
\label{sun-best-fit}
\end{equation}

The limits in Eq.~(\ref{LMA})
show that the mixing relevant for solar neutrino oscillations
is large.
However,
maximal mixing seems strongly disfavored
from the analysis of solar neutrino data
in Ref.~\cite{Bahcall:2002hv}.
This conclusion is supported by the results
of some other authors
\cite{Ahmad:2002ka,Barger:2002iv,deHolanda:2002pp,Fukuda:2002pe},
whereas the authors of
Refs.~\cite{Bandyopadhyay:2002xj,Strumia:2002rv,hep-ph/0206162,Fogli:2002pb}
found slightly larger allowed regions,
with maximal mixing marginally allowed.
Therefore,
it is not clear at present if maximal mixing
in solar neutrino oscillations is excluded or not.

In particular,
the authors of Refs.~\cite{hep-ph/0206162,Fogli:2002pb}
found a larger LMA region
and several allowed LOW+QVO and VAC regions,
where QVO indicates the region at
$\Delta{m}^2_{\mathrm{S}}
\sim
10^{-8} \, \mathrm{eV}^2
$
where both matter effect and vacuum oscillations are important.

The authors of Ref.~\cite{Strumia:2002rv}
found that taking into account only the most recent Gallium data
(taken in the period 1998-2001,
with improved systematic uncertainty
with respect to the earlier data)
the LOW region is less disfavored
with respect to the LMA region.

In any case,
there are good reasons to believe that
the SNO experiment has solved the
solar neutrino problem proving
its neutrino physics origin
and the presence of
$\nu_e \to \nu_\mu,\nu_\tau$
transitions,
and that
the current data indicate a large mixing.
The final confirmation
of the true allowed region
will hopefully come from the results of the KamLAND
\cite{KamLAND}
or BOREXINO
\cite{BOREXINO}
experiments.

\section{Atmospheric Neutrinos}
\label{Atmospheric Neutrinos}

In 1998
the Super-Kamiokande experiment found
model independent evidence of
disappearance of atmospheric $\nu_\mu$'s
\cite{Fukuda:1998mi},
which is supported
by the results of the
K2K long-baseline experiment \cite{K2K-Nu2002-brief},
of the Soudan 2 \cite{Allison:1999ms} and MACRO \cite{Ambrosio:2000qy}
atmospheric neutrino experiment.
The atmospheric neutrino data of the
Super-Kamiokande experiment are well fitted by
$\nu_\mu \to \nu_\tau$
transitions with large mixing
(fact~\ref{F-atm} above):
\begin{equation}
\begin{array}{l}
1.2 \times 10^{-3}
<
\Delta{m}^2_{\mathrm{A}} / \mathrm{eV}^2
<
5.0 \times 10^{-3}
\,,
\\
\sin^2 2\vartheta_{\mathrm{A}}
>
0.84
\,,
\end{array}
\label{ATM}
\end{equation}
at 99\% C.L.
\cite{SK-atm-Nu2002-brief},
where
$\Delta{m}^2_{\mathrm{A}}$
is the relevant neutrino squared-mass difference
and
$\vartheta_{\mathrm{A}}$
is the effective mixing angle in two-generation analyses
of atmospheric neutrino data.
The best fit is at
\begin{equation}
\sin^2 2 \vartheta_{\mathrm{A}}
=
1.0
\,,
\
\Delta{m}^2_{\mathrm{A}}
\simeq
2.5 \times 10^{-3} \, \mathrm{eV}^2
\,.
\label{atm-best-fit}
\end{equation}
Furthermore,
other oscillation channels
($ \nu_\mu \leftrightarrows \nu_e $
or
$ \nu_\mu \to \nu_s $,
where $\nu_s$ is a sterile neutrino)
as well as other mechanisms
(as $\nu$ decay)
are disfavored.

In the future,
the K2K experiment will improve its results taking more data
and the MINOS \cite{MINOS} long-baseline experiment
will measure with
better accuracy the mixing parameters in Eq.~(\ref{ATM}).
The long-baseline experiment
ICARUS \cite{ICARUS}
and
OPERA \cite{OPERA}
will be aimed at direct observation of
$ \nu_\mu \to \nu_\tau $
transitions.

\section{Three-Neutrino Mixing}
\label{Three-Neutrino Mixing}

\begin{figure}
\begin{minipage}[t]{\linewidth}
\includegraphics[bb=194 467 441 775, clip, width=0.49\linewidth]{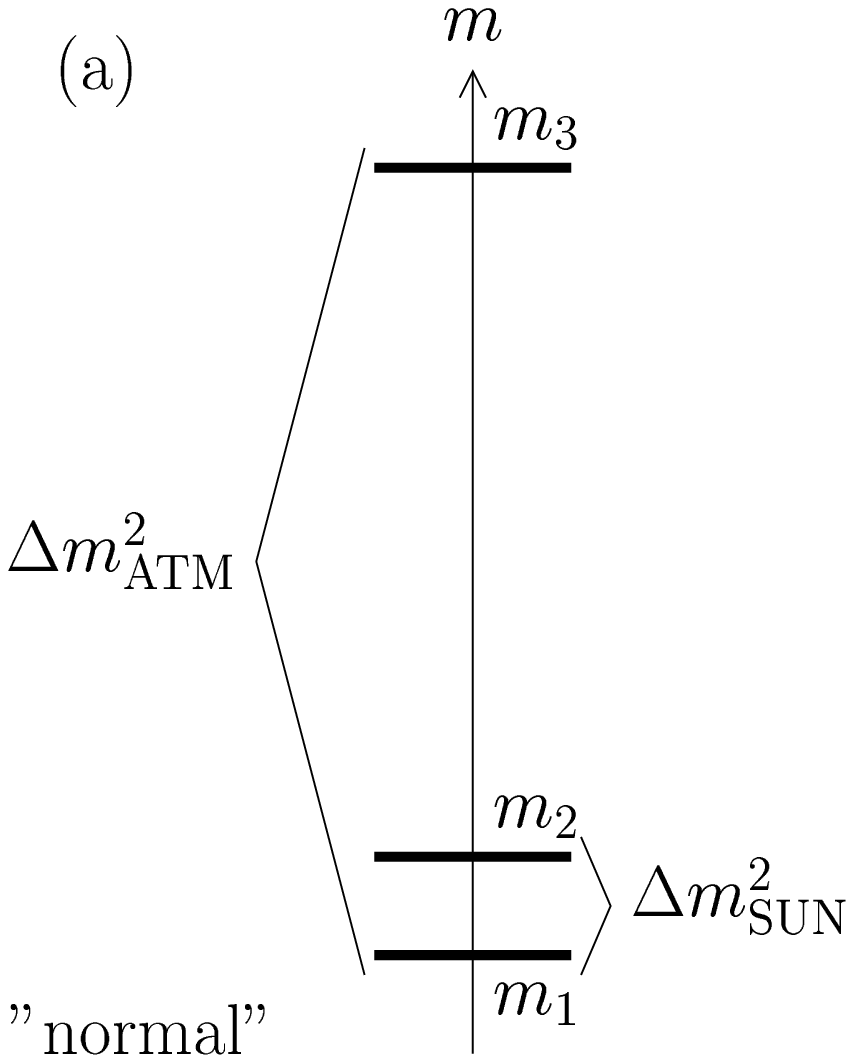}
\hfill
\includegraphics[bb=169 467 416 775, clip, width=0.49\linewidth]{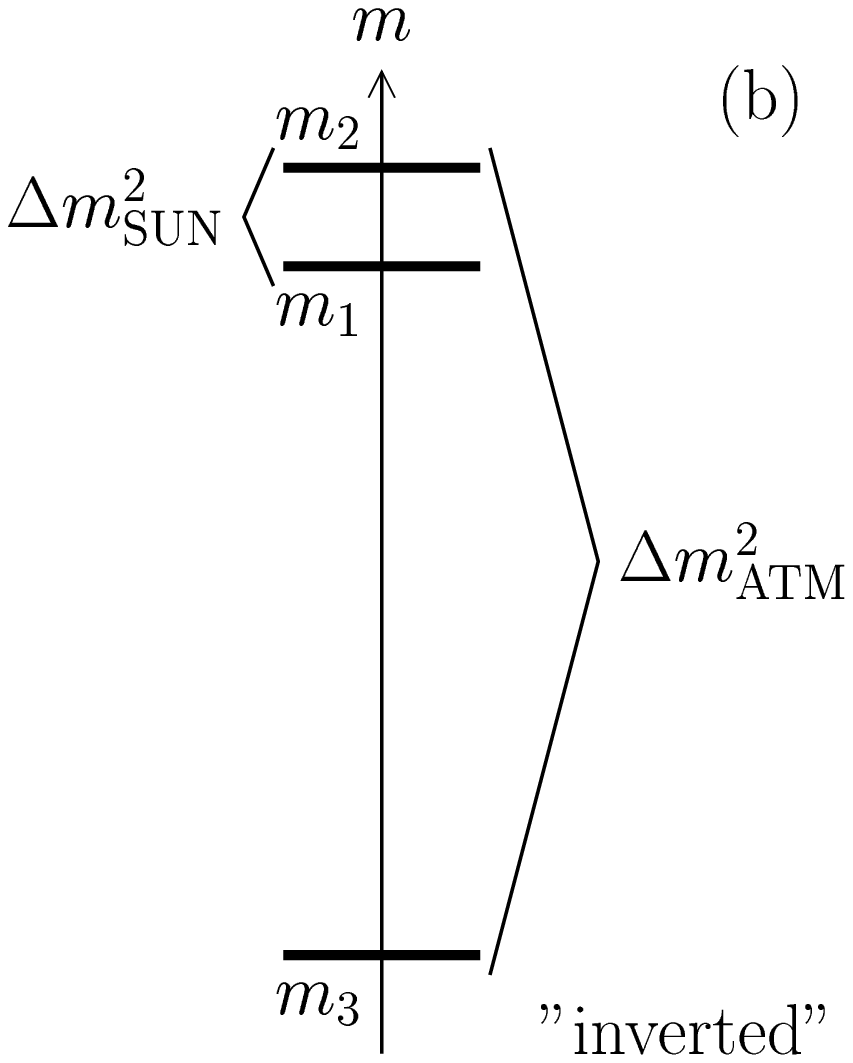}
\end{minipage}
\caption{ \label{3nu}
The two possible types of three-neutrino schemes.
}
\end{figure}

Solar and atmospheric neutrino data
can be well fitted
in the
framework of three-neutrino mixing
(see Ref.~\cite{BGG-review-98-brief})
with the hierarchy of squared-mass differences
\begin{equation}
\Delta{m}^2_{\mathrm{S}} = \Delta{m}^2_{21}
\ll
|\Delta{m}^2_{31}|
\simeq 
\Delta{m}^2_{\mathrm{A}}
\,,
\label{dm2hierarchy}
\end{equation}
where
$ \Delta{m}^2_{kj} \equiv m_k^2 - m_j^2 $
for $k,j=1,2,3$,
and
$m_1,m_2,m_3$
are the three neutrino masses.

Figure~\ref{3nu} shows the two possible types of
three-neutrino schemes:
a ``normal'' type (a)
and an ``inverted'' type (b).
The absolute scale of these schemes
(mystery~\ref{M-absolute-scale} above)
is not fixed by
neutrino oscillation experiments,
which can measure only squared-mass differences.
The normal three-neutrino scheme
allows the natural mass hierarchy
\begin{equation}
m_1 \ll m_2 \ll m_3
\,,
\label{hierarchy}
\end{equation}
whereas the inverted scheme
allows the so-called ``inverted hierarchy''
$m_3 \ll m_1 \simeq m_2 $.

The absolute scale of neutrino masses is bounded
from above by the results of Tritium
$\beta$-decay experiments,
which found
\cite{Weinheimer-Nu2002-brief}
\begin{equation}
m_{\nu_e}
<
2.2 \, \mathrm{eV}
\quad
\text{(95\% C.L.)}
\,.
\label{tritium}
\end{equation}
Since in the case of neutrino mixing
$\nu_e$ does not have a definite mass,
but is a superposition of massive neutrinos,
the limit in Eq.~(\ref{tritium})
is a bound on the masses of the main massive neutrino components
of $\nu_e$.
In the three-neutrino schemes in Fig.~\ref{3nu}
all the neutrino masses are constrained by the limit
in Eq.~(\ref{tritium}).
An independent confirmation of this limit
comes from the study of the formation
of large-scale structures in the Universe
\cite{Elgaroy:2002bi,astro-ph/0205223}.
In the future the KATRIN experiment \cite{Osipowicz:2001sq}
will probe the effective electron neutrino mass
in Tritium decay down to
about $0.3 \, \mathrm{eV}$.
Unfortunately,
in the case
of the natural scheme in Fig.~\ref{3nu}a
with the neutrino mass hierarchy (\ref{hierarchy}),
Eqs.~(\ref{ATM}) and (\ref{dm2hierarchy})
imply that
$ m_3 \lesssim 7 \times 10^{-2} \, \mathrm{eV} $,
which is beyond the reach of the KATRIN experiment.

Let us discuss now the
current information on the neutrino mixing matrix $U$.
Since the solar neutrino
transitions of $\nu_e$
in $\nu_\mu$ or $\nu_\tau$
cannot be experimentally distinguished
because the energy is well below the threshold for
charged-current reactions with production of
$\mu^-$ or $\tau^-$,
solar neutrino oscillations
depend only on the elements
$U_{e1}$,
$U_{e2}$,
$U_{e3}$
of the mixing matrix.
On the other hand,
the hierarchy (\ref{dm2hierarchy})
of squared-mass differences
in the two possible three-neutrino schemes in Fig.~\ref{3nu}
implies that atmospheric (and long-baseline) neutrino oscillations
depend only on the elements
$U_{e3}$,
$U_{\mu3}$,
$U_{\tau3}$
(see Ref.~\cite{BGG-review-98-brief}).
Since
$ \Delta{m}^2_{\mathrm{S}} = \Delta{m}^2_{21} $
and
$ \Delta{m}^2_{\mathrm{A}} \simeq |\Delta{m}^2_{31}| $
are independent,
the only quantity that connects
solar and atmospheric neutrino oscillations
is $U_{e3}$.
Therefore,
any information on the value of
$U_{e3}$
is of crucial importance.

From the results of the CHOOZ long-baseline
reactor neutrino experiment \cite{Apollonio:1999ae},
it is known
that the element $U_{e3}$
of the three-generation neutrino mixing matrix is small
\cite{Fogli:2002pb}:
\begin{equation}
|U_{e3}|^2 < 5 \times 10^{-2}
\quad
\text{(99.73\% C.L.)}
\,.
\label{Ue3bound}
\end{equation}
The results of the CHOOZ experiment have been confirmed
by the Palo Verde experiment \cite{Boehm:2001ik},
and by the absence of $\nu_e$ transitions in the
Super-Kamiokande atmospheric neutrino data
\cite{SK-atm-Nu2002-brief}.

An important consequence of the smallness of $U_{e3}$
is the practical decoupling of solar and atmospheric
neutrino oscillations
\cite{Bilenky:1998tw},
which can be analyzed in terms of two-neutrino oscillations
with the effective mixing angles
$\vartheta_{\mathrm{S}}$
and
$\vartheta_{\mathrm{A}}$
given by
\begin{equation}
\sin^2\vartheta_{\mathrm{S}}
=
\frac{|U_{e2}|^2}{1-|U_{e3}|^2}
\,,
\
\sin^2\vartheta_{\mathrm{A}}
=
\frac{|U_{\mu3}|^2}{1-|U_{e3}|^2}
\,.
\label{ss}
\end{equation}
Neglecting a possible small value of $|U_{e3}|$,
the mixing matrix can be written as
\begin{equation}
U
\simeq
\begin{pmatrix}
c_{\vartheta_{\mathrm{S}}} & s_{\vartheta_{\mathrm{S}}} & 0
\\
- s_{\vartheta_{\mathrm{S}}} c_{\vartheta_{\mathrm{A}}} & c_{\vartheta_{\mathrm{S}}} c_{\vartheta_{\mathrm{A}}} & s_{\vartheta_{\mathrm{A}}}
\\
s_{\vartheta_{\mathrm{S}}} s_{\vartheta_{\mathrm{A}}} & - c_{\vartheta_{\mathrm{S}}} s_{\vartheta_{\mathrm{A}}} & c_{\vartheta_{\mathrm{A}}}
\end{pmatrix}
\,,
\label{U01}
\end{equation}
with
$ c_{\vartheta} \equiv \cos \vartheta $
and
$ s_{\vartheta} \equiv \sin \vartheta $.
In this case,
solar neutrino transitions occur from
$
\nu_e = c_{\vartheta_{\mathrm{S}}} \nu_1 + s_{\vartheta_{\mathrm{S}}} \nu_2
$
to the orthogonal state
$
- s_{\vartheta_{\mathrm{S}}} \nu_1 + c_{\vartheta_{\mathrm{S}}} \nu_2
=
c_{\vartheta_{\mathrm{A}}} \nu_\mu - s_{\vartheta_{\mathrm{A}}} \nu_\tau
$.
Hence, the fractions of
$\nu_e\to\nu_\mu$
and
$\nu_e\to\nu_\tau$
transitions are determined by the value of the
atmospheric effective
mixing angle
$\vartheta_{\mathrm{A}}$.

From the experimental
best-fit value
$\vartheta_{\mathrm{A}} = \pi/4$
in Eq.~(\ref{atm-best-fit}),
we see that solar $\nu_e$ transform
approximately into
$ ( \nu_\mu - \nu_\tau ) / \sqrt{2} $,
which is an equal superposition
of $\nu_\mu$ and $\nu_\tau$.
Since the suppression factor of solar $\nu_e$'s
measured in the charged-current SNO reaction
is about $1/3$
\cite{Ahmad:2002jz},
the flux of $\nu_e$, $\nu_\mu$ and $\nu_\tau$
on Earth is approximately equal.
The mixing matrix is approximately given by
\begin{equation}
U
\approx
\begin{pmatrix}
c_{\vartheta_{\mathrm{S}}} & s_{\vartheta_{\mathrm{S}}} & 0
\\
- s_{\vartheta_{\mathrm{S}}} / \sqrt{2} & c_{\vartheta_{\mathrm{S}}} / \sqrt{2} & 1 / \sqrt{2}
\\
s_{\vartheta_{\mathrm{S}}} / \sqrt{2} & - c_{\vartheta_{\mathrm{S}}} / \sqrt{2} & 1 / \sqrt{2}
\end{pmatrix}
\,.
\label{U02}
\end{equation}

Let us now consider the best fit value in Eq.~(\ref{sun-best-fit})
of the effective solar mixing angle,
which implies that
$ \vartheta_{\mathrm{S}} \simeq \pi/6 $,
leading to the approximate bilarge mixing matrix
\begin{equation}
U
\approx
\begin{pmatrix}
\sqrt{3}/2 & 1/2 & 0
\\
- 1/2\sqrt{2} & \sqrt{3}/2\sqrt{2} & 1/\sqrt{2}
\\
1/2\sqrt{2} & - \sqrt{3}/2\sqrt{2} & 1/\sqrt{2}
\end{pmatrix}
\,.
\label{U03}
\end{equation}

However,
it is widely hoped that
$|U_{e3}|^2$ is not much smaller
than the current upper limit
in Eq.~(\ref{Ue3bound}),
because a non-vanishing value of $U_{e3}$
is essential for the possibility
of measuring CP violation in neutrino oscillations
(mystery~\ref{M-CP} above, see Ref.~\cite{Albright-0008064})
and
for the measurement of the sign of
$\Delta{m}^2_{31}$,
which distinguishes the normal and inverted schemes in Fig.~\ref{3nu}.
The sign of
$\Delta{m}^2_{31}$
can be measured
in long-baseline neutrino experiments
through matter effects in the Earth,
which need the participation to the oscillations
of $\nu_e$ through $U_{e3}$
($\nu_\mu$ and $\nu_\tau$
have the same neutral-current interaction with matter,
whereas $\nu_e$ has also charged-current interactions;
neutrino oscillations are affected
only by the difference of interactions of different
flavors).

The matrix in Eq.~(\ref{U03})
can be considered as the approximate current best-fit
mixing matrix.
The limits on
$\vartheta_{\mathrm{S}}$,
$\vartheta_{\mathrm{A}}$
and
$U_{e3}$
in
Eqs.~(\ref{LMA}),
(\ref{ATM})
and
(\ref{Ue3bound}),
respectively,
allow to derive the following allowed intervals
for the absolute values of the elements
of the mixing matrix
(the intervals are correlated,
because of unitarity):
\begin{equation}
|U|
\approx
\begin{pmatrix}
0.71-0.90 & 0.43-0.69 & 0.00-0.22 \\
0.24-0.66 & 0.40-0.81 & 0.53-0.84 \\
0.24-0.66 & 0.40-0.81 & 0.53-0.84 \\
\end{pmatrix}
\,.
\label{010}
\end{equation}

The best known way to investigate the Majorana nature
of neutrinos
(mystery~\ref{M-nature} above)
is the search for neutrinoless double-$\beta$ decay,
whose amplitude is proportional to the
effective Majorana mass
\begin{equation}
|\langle m \rangle|
=
| \sum_k U_{ek}^2 \, m_k |
\,,
\label{m}
\end{equation}
for which the current experimental
upper limit is
between about 0.3 and 1 eV,
taking into account a
theoretical uncertainty of about a factor 3
in the calculation of the nuclear matrix element
(see Ref.~\cite{Vogel-0202264}).
In general,
since the elements $U_{ek}$ of the mixing matrix are
complex,
cancellations among the contributions of the three massive
neutrinos are possible.
The current allowed interval for
$|\langle m \rangle|$
in the case of the two three-neutrino mixing schemes
in Fig.~\ref{3nu}
can be found in Ref.~\cite{hep-ph/0205022}.

Here let me consider the case
of the natural scheme in Fig.~\ref{3nu}a
with the neutrino mass hierarchy (\ref{hierarchy}).
In this case,
the limits in Eqs.~(\ref{LMA}) and (\ref{ATM})
imply that the
contribution of $m_1$ to $|\langle m \rangle|$
is negligible
and the absolute values of the contributions
of $m_2$ and $m_3$ are limited by
\begin{align}
\null & \null
1.5 \times 10^{-3} \, \mathrm{eV}
\lesssim
| U_{e2}^2 \, m_2 |
\lesssim
9.5 \times 10^{-3} \, \mathrm{eV}
\,,
\label{m2}
\\
\null & \null
| U_{e3}^2 \, m_3 |
\lesssim
3.5 \times 10^{-3} \, \mathrm{eV}
\,.
\label{m3}
\end{align}
Therefore,
it is possible that the dominant
contribution to $|\langle m \rangle|$
comes from $m_2$,
and not from the largest mass $m_3$.
If
$ | U_{e3}^2 \, m_3 | \ll | U_{e2}^2 \, m_2 | $,
there are no cancellations in Eq.~(\ref{m})
\cite{Giunti:1999jw}
and the effective Majorana mass is expected in the interval
in Eq.~(\ref{m2}).
Unfortunately,
such values of $|\langle m \rangle|$
are beyond the reach of proposed experiments
(see Ref.~\cite{Vogel-0202264}).

\section{Conclusions}
\label{Conclusions}

The results of all neutrino experiments,
except LSND,
are in agreement with the hypothesis of three-neutrino mixing.
If future data will confirm this scenario
the most important phenomenological task
will be to refine the measurement of the
neutrino mixing parameters,
especially the measurement of $U_{e3}$,
whose value is related to the possibility
of distinguishing the two schemes
in Fig.~\ref{3nu}
and of measuring CP violation
in neutrino oscillations.
The measurements of the absolute scale of neutrino masses
and of the effective Majorana mass
in neutrinoless double-$\beta$ decay
seem to be more difficult tasks
(beyond the reach of existing projects
if the neutrino masses satisfy the
natural hierarchy (\ref{hierarchy})).

\end{document}